\documentclass[reqno]{amsart}
\usepackage{graphicx}
\usepackage{textcomp}
\usepackage{gensymb}
\usepackage{xcolor}

\begin{document}

\title[Low cloud cover]{No experimental evidence for the significant anthropogenic climate change}%
\author{J. Kauppinen and P. Malmi}%
\address{Department of Physics and Astronomy, University of Turku}%
\email{jyrkau@utu.fi}

%\thanks{thanks}%
%\subjclass{subjclass}%
%\keywords{keywords}%

\date{\today}%
%\dedicatory{}%
%\commby{}%
% ----------------------------------------------------------------
\begin{abstract}
In this paper we will prove that GCM-models used in IPCC report AR5 fail to calculate the influences of the low cloud cover changes on the global temperature. That is why those models give a very small natural temperature change leaving a very large change for the contribution of the green house gases in the observed temperature. This is the reason why IPCC has to use a very large sensitivity to compensate a too small natural component. Further they have to leave out the strong negative feedback due to the clouds in order to magnify the sensitivity. In addition, this paper proves that the changes in the low cloud cover fraction practically control the global temperature.
\end{abstract}
\maketitle
% ----------------------------------------------------------------
\section{Introduction}
The climate sensitivity has an extremely large uncertainty in the scientific literature. The smallest values estimated are very close to zero while the highest ones are even 9 degrees Celsius for a doubling of CO$_2$. The majority of the papers are using theoretical general circulation models (GCM) for the estimation. These models give very big sensitivities with a very large uncertainty range. Typically sensitivity values are between 2--5 degrees. IPCC uses these papers to estimate the global temperature anomalies and the climate sensitivity. However, there are a lot of papers, where sensitivities lower than one degree are estimated without using GCM. The basic problem is still a missing experimental evidence of the climate sensitivity. One of the authors (JK) worked as an expert reviewer of IPCC AR5 report. One of his comments concerned the missing experimental evidence for the very large sensitivity presented in the report \cite{AR5}. As a response to the comment IPCC claims that an observational evidence exists for example in Technical Summary of the report. In this paper we will study the case carefully.

\section{Low cloud cover controls practically the global temperature}

\begin{figure}
  \centering
  \includegraphics[width=\linewidth]{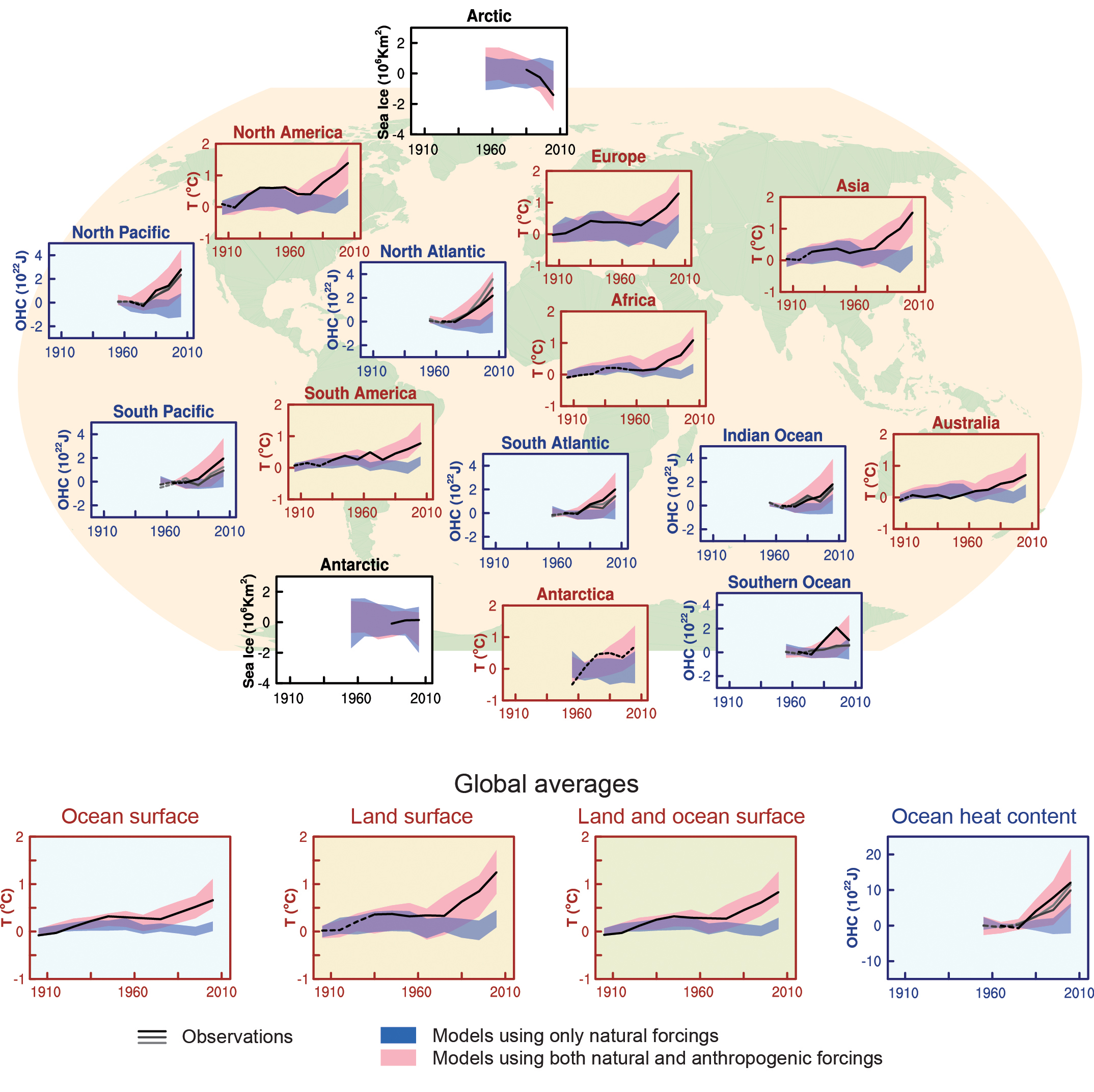}
  \caption{Figure TS.12 on page 74 of the Technical Summary of the IPCC Fifth Assessment report (AR5).}\label{TS12}
\end{figure}

The basic task is to divide the observed global temperature anomaly into two parts: the natural component and the part due to the green house gases. In order to study the response we have to re-present Figure TS.12 from Technical Summary of IPCC AR5 report (1). This  figure is Figure 1. Here we highlight the subfigure ``Land and ocean surface'' in Figure \ref{TS12}. Only the black curve is an observed temperature anomaly in that figure. The red and blue envelopes are computed using climate models. We do not consider computational results as experimental evidence. Especially the results obtained by climate models are questionable because the results are conflicting with each other.
\begin{figure}
  \centering
  \includegraphics[width=\linewidth]{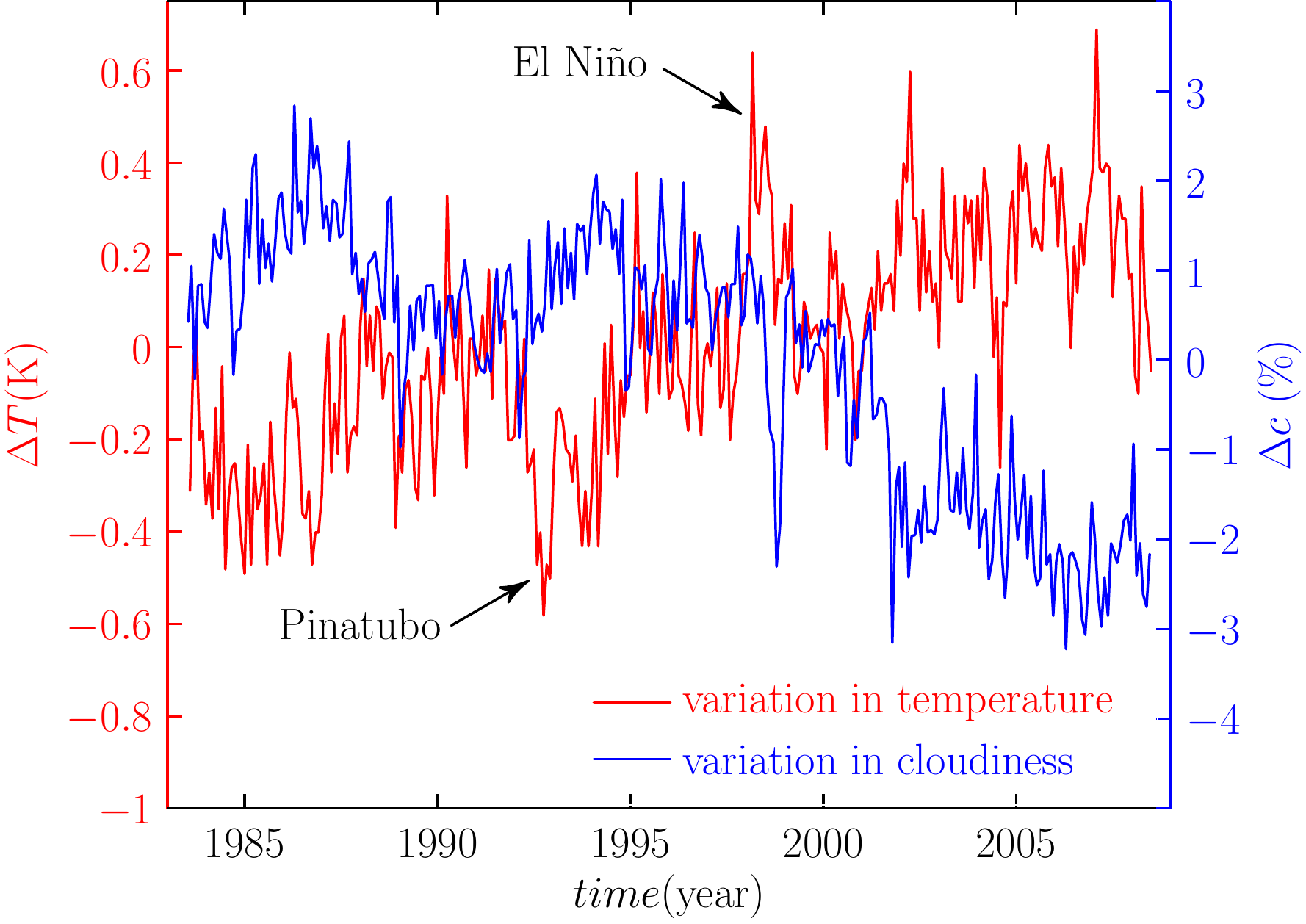}
  \caption{\cite{Kau2014} Global temperature anomaly (red) and the global low cloud cover changes (blue) according to the observations. The anomalies are between summer 1983 and summer 2008. The time resolution of the data is one month, but the seasonal signal is removed. Zero corresponds about 15\degree C for the temperature and 26~\% for the low cloud cover.}\label{TempCloud}
\end{figure}

\begin{figure}
  \centering
  \includegraphics[width=\linewidth]{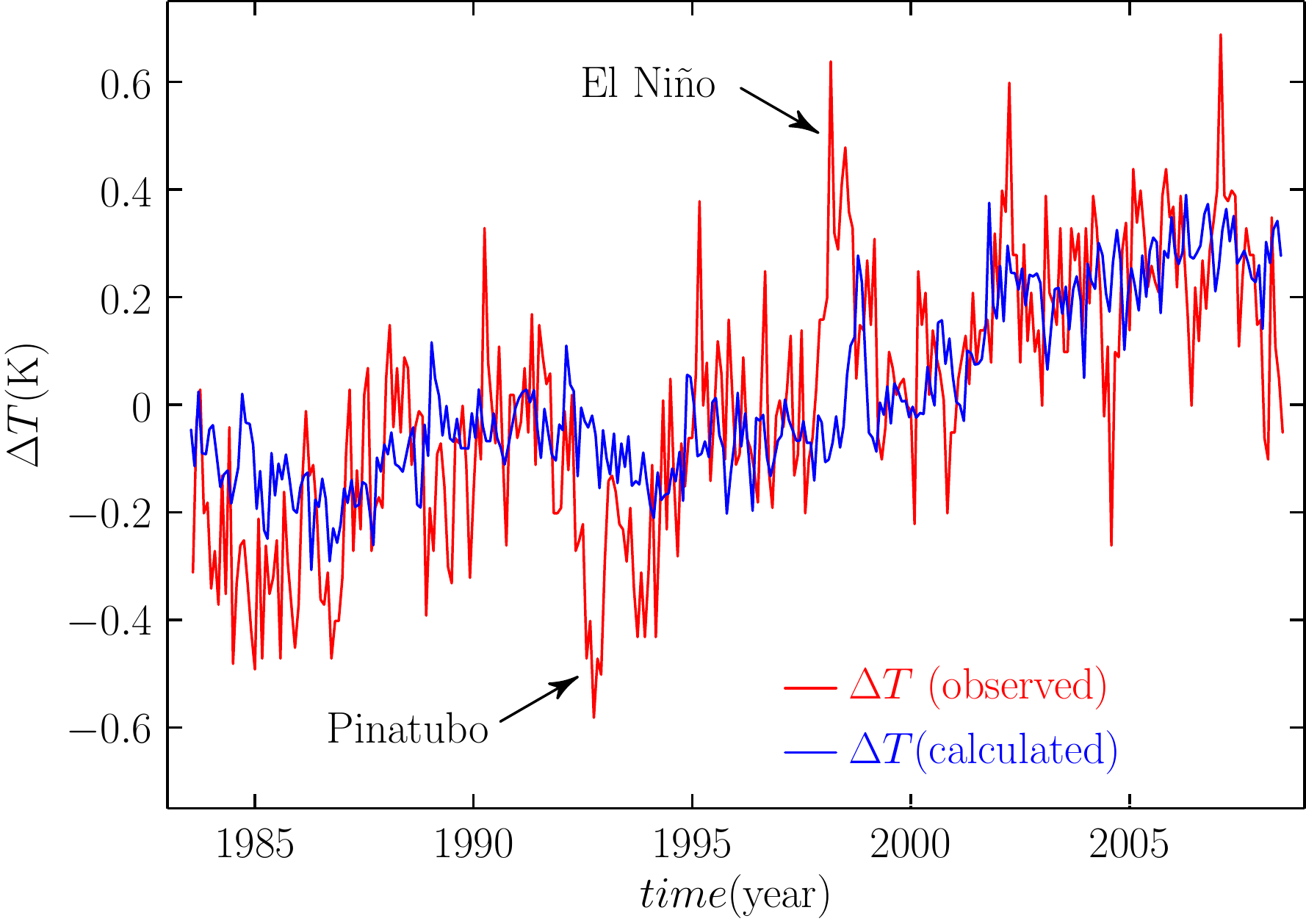}
  \caption{\cite{Kau2014} Global natural temperature anomaly (blue) superimposed on the observed (red) temperature anomaly. The blue anomaly is derived using the observed low cloud cover data from Figure \ref{TempCloud}. There are half a dozen very sharp ghost spikes in the observed (red) temperature anomaly. The Pinatubo eruption and the strong El Ni\~{n}o are clearly seen.}\label{TempObsCalc}
\end{figure}

In Figure \ref{TempCloud} we see the observed global temperature anomaly (red) and global low cloud cover changes (blue). These experimental observations indicate that 1~\% increase of the low cloud cover fraction decreases the temperature by $0.11\degree$C. This number is in very good agreement with the theory given in the papers \cite{Kau2011,Kau2014,Kau2018}. Using this result we are able to present the natural temperature anomaly by multiplying the changes of the low cloud cover by $-0.11\degree {\rm C}/\%$. This natural contribution (blue) is shown in Figure \ref{TempObsCalc} superimposed on the observed temperature anomaly (red). As we can see there is no room for the contribution of greenhouse gases i.e. anthropogenic forcing within this experimental accuracy. Even though the monthly temperature anomaly is very noisy it is easy to notice a couple of decreasing periods in the increasing trend of the temperature. This behavior cannot be explained by the monotonically increasing concentration of CO$_2$ and it seems to be far beyond the accuracy of the climate models.

The red curve in Figures \ref{TempCloud} and \ref{TempObsCalc} corresponds to the black curve, between years 1983 and 2008, in the above-mentioned subfigure ``Land and ocean surface''. If the clouds and CO$_2$ were taken into account correctly in the climate models both the blue and red envelopes should overlap the observed black curve. As we see the trend of the blue envelope is more like decreasing. We suggest this is due to a wrong or missing processing of the low cloud cover contribution. In the report AR5 it is even recognized that the low clouds give the largest uncertainty in computation. In spite of this IPCC still assumes that the difference between the blue and red envelopes in Figure \ref{TS12} is the contribution of greenhouse gases.

Unfortunately, the time interval (1983--2008) in Fig \ref{TempCloud} is limited to 25 years because of the lack of the low cloud cover data. During this time period the CO$_2$ concentration increased from 343~ppm to 386~ppm and both Figures \ref{TS12} (IPCC) and \ref{TempCloud} show the observed temperature increase of about $0.4\degree$C. The actual global temperature change, when the concentration of CO$_2$ raises from $C_0$ to $C$, is
\begin{equation}\label{DeltaT}
  \Delta T = \frac{\Delta T_{\rm 2CO_2}\ln C/C_0}{\ln 2} - 11\degree{\rm C}\cdot \Delta c,
\end{equation}
where $\Delta T_{\rm 2CO_2}$ is the global temperature change, when the CO$_2$ concentration is doubled and $\Delta c$ is the change of the low cloud cover fraction. The first and second term are the contributions of CO$_2$ \cite{Myh} and the low clouds, respectively. Using the sensitivity $\Delta T_{\rm 2CO_2}=0.24\degree$C derived in the papers \cite{Kau2011,Kau2014,Kau2018} the contribution of greenhouse gases to the temperature is only about $0.04\degree$C according to the first term in the above equation. This is the reason why we do not see this small increase in temperature in Figure \ref{TempObsCalc}, where the temperature anomaly is quite noisy with one month time resolution. It is clearly seen in Figure \ref{TempCloud} that the red and blue anomalies are like mirror images. This means that the first term is much smaller than the absolute value of the second term ($11\degree {\rm C}\cdot \Delta c$) in equation (\ref{DeltaT}).

It turns out that the changes in the relative humidity and in the low cloud cover depend on each other \cite{Kau2018}. So, instead of low cloud cover we can use the changes of the relative humidity in order to derive the natural temperature anomaly. According to the observations 1~\% increase of the relative humidity decreases the temperature by $0.15\degree$C, and consequently the last term in the above equation can be approximated by $-15\degree{\rm C} \Delta\phi$, where $\Delta\phi$ is the change of the relative humidity at the altitude of the low clouds.

\begin{figure}
  \centering
  \includegraphics[width=\linewidth]{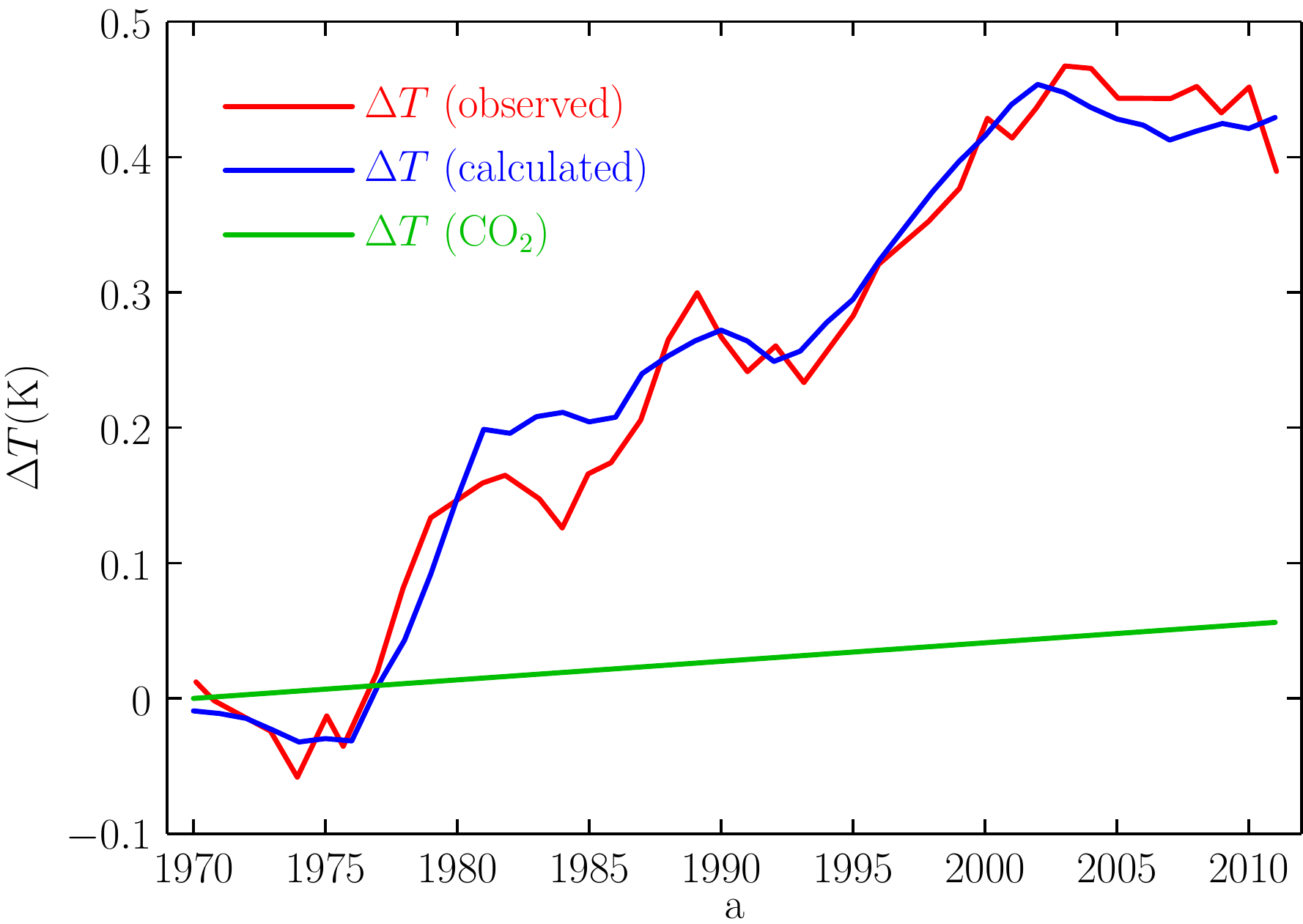}
  \caption{\cite{Kau2014} Observed global mean temperature anomaly (red), calculated anomaly (blue), which is the sum of the natural and carbon dioxide contributions. The green line is the CO2 contribution merely. The natural component is derived using the observed changes in the relative humidity. The time resolution is one year.}\label{TempObsCalcJapan}
\end{figure}

Figure \ref{TempObsCalcJapan} shows the sum of the temperature changes due to the natural and CO$_2$ contributions compared with the observed temperature anomaly. The natural component has been calculated using the changes of the relative humidity. Now we see that the natural forcing does not explain fully the observed temperature anomaly. So we have to add the contribution of CO$_2$ (green line), because the time interval is now 40 years (1970--2010). The concentration of CO$_2$ has now increased from 326~ppm to 389~ppm. The green line has been calculated using the sensitivity $0.24\degree$C, which seems to be correct. In Fig. \ref{TempObsCalcJapan} we see clearly how well a change in the relative humidity can model the strong temperature minimum around the year 1975. This is impossible to interpret by CO$_2$ concentration.

The IPCC climate sensitivity is about one order of magnitude too high, because a strong negative feedback of the clouds is missing in climate models. If we pay attention to the fact that only a small part of the increased CO$_2$ concentration is anthropogenic, we have to recognize that the anthropogenic climate change does not exist in practice. The major part of the extra CO$_2$ is emitted from oceans \cite{Kau2019}, according to Henry`s law. The low clouds practically control the global average temperature. During the last hundred years the temperature is increased about $0.1\degree$C because of CO$_2$. The human contribution was about $0.01\degree$C.

\section{Conclusion}

We have proven that the GCM-models used in IPCC report AR5 cannot compute correctly the natural component included in the observed global temperature. The reason is that the models fail to derive the influences of low cloud cover fraction on the global temperature. A too small natural component results in a too large portion for the contribution of the greenhouse gases like carbon dioxide. That is why IPCC represents the climate sensitivity more than one order of magnitude larger than our sensitivity $0.24\degree$C. Because the anthropogenic portion in the increased CO$_2$ is less than 10~\%, we have practically no anthropogenic climate change. The low clouds control mainly the global temperature.

%\pagebreak
\bibliographystyle{unsrt}
\bibliography{climate}

\end{document}